\documentstyle[eqsecnum,epsf,aps,prb,multicol]{revtex}
\newcommand{\bleq}{\ifpreprintsty
                   \else
                   \end{multicols}\vspace*{-3.5ex}{\tiny
                   \noindent\begin{tabular}[t]{c|}
                   \parbox{0.493\hsize}{~} \\ \hline \end{tabular}}
                   \fi}
\newcommand{\eleq}{\ifpreprintsty
                   \else
                   {\tiny\hspace*{\fill}\begin{tabular}[t]{|c}\hline
                    \parbox{0.49\hsize}{~} \\
                    \end{tabular}}\vspace*{-2.5ex}\begin{multicols}{2}
                    \noindent
                    \fi}
\newcommand{\bcols}{\ifpreprintsty\else\begin{multicols}{2}\fi}
\newcommand{\ecols}{\ifpreprintsty\else\end{multicols}\fi}
\def\d{{\rm d}}
\def\lan{\left\langle}
\def\ran{\right\rangle}
\def\dl{{{\rm d} \over {{\rm d} \ell}} \,}
\def\dx{{\rm d}x}
\def\e{{\rm e}}
\def\virg{\;\;,}

\def\point{\;\,.}
\def\vf{v_{\rm F}}
\def\ef{\epsilon_{\rm F}}
\def\kf{k_{\rm F}}
\def\al{\alpha }
\def\para{\parallel }
\def\thetatilde{\widetilde{\theta}}
\def\phitilde{\widetilde{\phi}}

\def\N{\widehat{N}}

\def\gpara{g_{1^{\parallel}}}
\def\gperp{g_{1^{\perp}}}
\def\g2{g_{2}}
\def\tn{t/(\vf \alpha^{-1})}
\def\tt{\tilde{t}}

\def\ggs{\buildrel\textstyle > \over {\hbox{\raise0.2ex\hbox{$\sim$}}}}
\def\lls{\buildrel\textstyle < \over {\hbox{\raise0.2ex\hbox{$\sim$}}}}
\def\gsim{\,\lower0.75ex\hbox{$\ggs$}\,}
\def\lsim{\,\lower0.75ex\hbox{$\lls$}\,}
\def\jo #1#2#3#4{#1 #2 (#3)  #4}   
\def\PR{Phys.\ Rev.}
\def\PRB{Phys.\ Rev.\ B}

\def\JPF{J.\ Phys.\ France}

\def\JPSJ{J.\ Phys.\ Soc.\ Jpn.}

\def\RMP{Rev.\ Mod.\ Phys.}
\def\PTP{Prog.\ Theor.\ Phys.}

\def\ADV{Adv.\ Phys.}

\def\SL{JETP\ Lett.}

\def\PHC{Physica C}

\def\PL{Phys.\ Lett.}
\def\JLP{J.\ Low Temp.\ Phys.}

\def\MCLC{Mol.\ Cryst.\ Liq.\ Cryst.}
\begin{document} 
\draft

\title
{
Two-Coupled  Chains   with Spin-Anisotropic  Backward Scattering
}
\author{
Masahisa Tsuchiizu and Yoshikazu Suzumura
}
\address{
Department of Physics, Nagoya University, Nagoya 464-8602, Japan \\
}

\date{Received 16 March 1998; revised manuscript received 16 April 1998}
\maketitle

\begin{abstract}
By applying renormalization group method  to the bosonized Hamiltonian 
of two-coupled chains with repulsive intrachain interaction,
 we have examined a role of backward scattering with a spin-anisotropy
  which competes  with interchain hopping.
From  calculation of a dominant state in the limit of low 
 energy, it is found that 
   superconducting state  moves into 
    spin density wave state when the  anisotropy 
  becomes larger than a critical value.
Further phase diagram is shown on the plane of $g$-ology.
\end{abstract}


\vspace{1cm}
\bcols

\section{Introduction}
In low dimensional electron systems, which  consist of 
 an array of one-dimensional chains  with repulsive interaction,  
  a noticeable phase diagram has been found 
  where spin density wave (SDW) state competes with superconducting 
  (SC) state \cite{Jerome}.
 Among these materials, the organic conductors, (TMTSF)$_2 X$ salts, 
 have shown the SC state with $d$-wave like paring
\cite{Takigawa,Hasegawa}. 
 Such a SC state is located close to the SDW state 
 which  exhibits 
  one-dimensional fluctuation 
 originating in  intrachain interaction and 
 a small interchain electron hopping
\cite{Bourbonnais_JPF}. 
 In addition to phenomena arising from
  the  repulsive intrachain interaction,
  these materials exhibit anisotropy of 
  spin susceptibility in the SDW state  
 with the easy axis ($b$-direction) 
 perpendicular to the conducting chain 
 although anisotropy  is invisible above the SDW 
transition temperature
\cite{Gruner}.

For studying the competition between the SC state and the SDW state, 
  we examine two-coupled chains with spin-anisotropic interaction. 
 The electronic state which is obtained in a single chain with repulsive 
interaction \cite{Solyom}
 is quite different from 
 those of   quasi-one-dimensional system  of  these conductors 
in which  the SC state shows $d$-wave like pairing 
denoting electrons between chains. 
 Therefore, we need at least two chains treated correctly  
although infinite number of chains gives rise to 
 a  phase transition at finite temperatures.
Actually such a  SC state in case of weak coupling of Hubbard model has  been obtained 
not only for two-coupled chains but also for  three-coupled chains 
 and speculated for arbitrary number of coupled-chains
\cite{Schulz}. 
 It seems that the dominant state in  two-coupled chains 
is relevant to  electronic states for the above systems 
 if we  note that electronic states of 
quasi-one-dimensional conductors  may be obtained as a result 
 of  developing these short range correlations perpendicular to chain.  

 Two-coupled chains with spin-isotropic interaction have been studied 
 extensively. 
In case of forward scattering described by Tomonaga-Luttinger model,
  the ground state given by the density wave is the same as 
  that of a single chain except for  removing degeneracy of 
  in-phase and out-of-phase pairings
  \cite{Kusmartsev,Yakovenko,Nersesyan,Finkelstein,Fabrizio,Yoshioka_JPSJ95,Yoshioka_JLTP}.
  On the other hand, 
 two-chains of  Hubbard model  with repulsive backward scattering 
 leads to  the ground state  given by  the $d$-wave like 
 SC state \cite{Fabrizio,Balents,Schulz},  
 while  that of  a single chain is  the SDW state.  
 Numerical studies for the Hubbard model also show an enhancement 
  of SC state for  some ranges of parameters of 
 interaction, interchain hopping   and filling \cite{Yamaji,Aoki}.
 Analysis in terms of the bosonization indicates the fact that 
 such a SC state appears due to  formation of
 a gap   of the total spin fluctuation,  
 which is triggered by the relevance of 
 transverse charge fluctuation \cite{Schulz,Yoshioka_PRB}. 
 As for the spin-anisotropic interaction, 
 the one-dimensional model has been explored  to study 
 the SDW state of above organic conductors
\cite{Giamarchi_JPF}.

In the present paper, the effect of spin-anisotropic 
 backward scattering  on the electronic state of two-coupled chains 
 is examined by improving  the previous  method \cite{Yoshioka_PRB}
 of the renormalization 
 group, which is based on the  bosonization.
In section 2, formulation  for renormalization group equations 
is given  where the backward scattering with 
  parallel spins is distinguished from that with opposite spins to
examine the spin-anisotropic interaction. 
 In  section 3, 
 renormalization group equations are calculated numerically 
 and  the most dominant state in the limit of energy is examined 
 where 
 phase diagram of SC state and SDW state are shown on the plane of 
  coupling constants and interchain hopping.
Discussion is given in section 4. 
 
\section{Formulation} 

We consider the Hamiltonian for two-coupled  chains 
 given by 
\begin{eqnarray}           \label{H0}
{\cal H}
&=& \sum_{k,p,\sigma,i} \, 
    \epsilon_{k,p} a_{k,p,\sigma,i}^\dagger a_{k,p,\sigma,i}
\nonumber \\ && \hspace{0.5cm}
 -  t \sum_{k,p,\sigma} 
    \left[ 
    a_{k,p,\sigma,1}^\dagger a_{k,p,\sigma,2} + {\rm h.c.}
    \right] \nonumber \\
&+&\frac{1}{2L} \sum_{p,p',\sigma,\sigma ',i} \sum_{k_1,k_2,q} 
     ( g_{2} \, \delta_{p,p'} 
   +  \gpara \, \delta_{p,-p'} \delta_{\sigma,\sigma '}
\nonumber \\ && \hspace{4cm}
   +  \gperp \, \delta_{p,-p'} \delta_{\sigma,-\sigma '}) \,
\nonumber \\ && \hspace{0.cm} \times 
    a_{k_1,p,\sigma,i}^\dagger \, a_{k_2,-p,\sigma ',i}^\dagger \,
    a_{k_2+q,-p',\sigma ',i} \, a_{k_1-q,p',\sigma,i}
\point
\end{eqnarray}
 The quantity  $a_{k,p,\sigma,i}^\dagger $
 denotes a  creation  operator for  the electron
 with momentum $k$, spin 
  $\sigma(=\uparrow,\downarrow \,\,{\rm or}\,\, +,-)$
  where  $p=+(-)$ represents the right-going (left-going) state
  and  $i(=1,2)$ denotes chain index. 
In Eq. (\ref{H0}), $\epsilon_{k,p}(=\vf (pk-k_{\rm F}))$ is
  the kinetic  energy with Fermi velocity $\vf$ and 
   Fermi momentum $\kf$. The  energy, $t$, is 
  that of the interchain hopping.  
 Coupling constants $\gpara$ and $\gperp$ represent  
  the intrachain interaction 
   of the backward scattering with parallel and opposite spins 
 respectively 
   and $g_2$ is the coupling constant of the forward scattering.

 By use of  the unitary transformation given by 
$
c_{k,p,\sigma,\mu} 
= ( - \mu a_{k,p,\sigma,1} + a_{k,p,\sigma,2} )/\sqrt{2} 
$
 $(\mu=\pm)$, 
  the first and the second terms in Eq. (\ref{H0})
  are   diagonalized 
 resulting in   the separation of 
 the Fermi wave number,   $k_{{\rm F}\mu} \equiv \kf - \mu t/\vf$. 
 Based on the bosonization around the new Fermi point, 
 we introduce  the phase variables, 
  $\theta_+$ and  $\phi_+$
($\thetatilde _+$ and $\phitilde _+$),  
 expressing    fluctuations 
 of the total (transverse) charge density and 
  spin density  respectively. 
 These variables are defined by  
$
\sum_{q\neq0} 2\pi i/(qL) \, \e^{-\al |q| /2 -iqx} \, 
  c_{k+q,p,\sigma,\mu}^\dagger c_{k,p,\sigma,\mu}
= [\theta_+ + p\theta_- + \mu (\thetatilde_+ + p \thetatilde_-)
 + \sigma (\phi_+ + p \phi_-) + \sigma \mu (\phitilde_+ + p \phitilde_-)
] /(2\sqrt{2})
$
\cite{Yoshioka_JLTP}.
 They  satisfy the 
  relation,
$ 
 [\theta_{+}(x),\theta_-(x')]= [\phi_{+}(x),\phi_-(x')] 
 =[\tilde{\theta}_{+}(x),\tilde{\theta}_-(x')]
 =[\tilde{\phi}_{+}(x),\tilde{\phi}_-(x')]
 = i \pi {\rm sgn}(x-x')
 $ 
 where the suffix $-$ denotes the  canonically conjugate variable. 
 The field operator is also expressed as 
$
\psi_{p,\sigma,\mu}(x)  =  
 L^{-1/2}  \sum_k \e^{ikx} c_{k,p,\sigma,\mu} = 
1/\sqrt{2\pi \al} \exp \left( ipk_{{\rm F}\mu}x 
 + i\Theta _{p,\sigma,\mu} \right) 
   \exp \left( i\pi \Xi_{p,\sigma,\mu} \right)
$
where 
$
\Theta _{p,\sigma,\mu}
  =  
  [
     p \theta_+ + \theta_- 
   + \sigma ( p \phi _+ + \phi _- )
   + \mu    ( p \thetatilde _+ + \thetatilde _- )
   + \sigma \mu ( p \phitilde _+ + \phitilde _- )
                      ]/(2\sqrt{2})
 $ 
 and $\al$ is of the order of the lattice constant.  
   For keeping  the anticommutation relation,
 the phase factor, $\pi \Xi_{p,\sigma,\mu}$, 
    is chosen as 
 $
  \Xi_{2n+j} = \N_1 + \cdots +\N_{2n} 
    + (-1)^{j+1}(\N_{2n+1} + \N_{2n+2})/2 
 $ with $j$=1,2 and $n=0,1,2,3$.    
The index $i$ of  $\N_{i}$ is defined 
 by  $i =(p,\sigma,\mu)$ where 
 $(+,+,+)=1,(+,-,+)=2,(+,+,-)=3,
(+,-,-)=4,(-,+,+)=5,(-,-,+)=6,(-,+,-)=7$ 
 and $(-,-,-)=8$, respectively.  
 Such a choice of   $\Xi_{p,\sigma,\mu}$ 
 leads to  a conservation of 
  a sign of  interactions. 
Thus  Eq. (\ref{H0}) is rewritten as 
\bleq
\begin{eqnarray}  \label{phase_Hamiltonian}
{\cal H} 
&=& \frac{v_\theta}{4\pi} \int \dx 
    \biggl\{ 
        \frac{1}{K_\theta} \left( \partial \theta_+ \right)^2
      + K_\theta \left( \partial \theta_- \right)^2 
    \biggr\} 
\hspace{2mm} + \hspace{2mm}    
    \frac{v_\phi}{4\pi} \int \dx 
    \biggl\{
        \frac{1}{K_\phi} \left( \partial \phi_+ \right)^2
      + K_\phi \left( \partial \phi_- \right)^2
    \biggr\} \nonumber \\
&+& \frac{\vf}{4\pi} \int \dx 
    \biggl\{ 
        \frac{1}{K_{\thetatilde}}\left( \partial \thetatilde_+ \right)^2
      + K_{\thetatilde} \left( \partial \thetatilde _- \right)^2 
    \biggr\}  
\hspace{2mm} + \hspace{2mm} 
    \frac{\vf}{4\pi} \int \dx 
    \biggl\{
        \frac{1}{K_{\phitilde}} \left( \partial \phitilde_+ \right)^2
      + K_{\phitilde} \left( \partial \phitilde _- \right)^2
    \biggr\} \nonumber \\
 &+&   \frac{1}{2\pi^2 \al^2} \int \dx \Bigl[
     \, g_a \cos \left( \sqrt{2}\thetatilde _+  - 4tx/\vf \right) 
    \cos \sqrt{2} \phitilde_+ 
   +  \, g_b \cos \left( \sqrt{2}\thetatilde _+ - 4tx/\vf \right) 
    \cos \sqrt{2} \phitilde_-   \nonumber \\
  & & \hspace{2cm}   
    +   \, g_c \cos \sqrt{2} \thetatilde_- \cos \sqrt{2} \phitilde_+ 
    +   \, g_d \cos \sqrt{2} \thetatilde_- \cos \sqrt{2} \phitilde_- 
\nonumber \\
 & & \hspace{2cm}  
   +    \, g_e  \cos \sqrt{2} \phi_+ 
           \cos \left( \sqrt{2}\thetatilde _+ - 4tx/\vf \right) 
   +    \, g_f  \cos \sqrt{2} \phi_+ \cos \sqrt{2} \thetatilde_- 
                           \nonumber \\
 & & \hspace{3cm} 
   +    \, g_g  \cos \sqrt{2} \phi_+ \cos \sqrt{2} \phitilde_+ 
   +    \, g_h  \cos \sqrt{2} \phi_+    \cos \sqrt{2} \phitilde_-  
                                  \Bigr]   \virg 
\end{eqnarray}
\eleq
where
$v_\theta=\vf \sqrt{1-\{(2g_2-\gpara)/2\pi\vf\}^2}$, 
$v_\phi = \vf \sqrt{1-\{\gpara/2\pi\vf\}^2}$, 
$K_\theta = [ \{1  -(2g_2 - \gpara)/2\pi\vf\}
      /\{1 + (2g_2 - \gpara)/2\pi\vf\} ]^{1/2}$,
$K_\phi = [ \{1  + \gpara/2\pi\vf\}
      /\{1 - \gpara/2\pi\vf\} ]^{1/2}$
 and 
$K_{\tilde{\theta}} = K_{\tilde{\phi}} = 1$.
 In  Eq. (\ref{phase_Hamiltonian}),  
 coefficients are given by 
$
g_a =   g_2 - \gpara
$,
$
g_b = - g_2
$,
$
g_c =  g_2
$,
$
g_d = - g_2 + \gpara 
$
 and 
$
g_e = -g_f = -g_g = -g_h = \gperp$ respectively. 


We examine the case where $\gpara = \g2$ but $\gperp$ is distinguished
  from $\gpara$ to examine the spin anisotropic backward scattering. 
 By assuming the scaling invariance 
  with respect to $\alpha \to \alpha '=\alpha \e^{\d l}$, 
 renormalization group method is applied to  response functions
\cite{Giamarchi_JPF,Giamarchi_PRB,Tsuchiizu} 
 defined by  
$
R_{A}(x_1-x_2,\tau _1-\tau _2) 
\equiv 
\lan T_\tau O_A(x_1,\tau_1) \, O_A^\dagger (x_2,\tau_2) \ran 
$
  where $\tau_j$ is the imaginary time and $O_A$ denotes operator 
 for order parameter of  SDW and  SC  states. 
In order to calculate second order renormalization group equations
 for all the coupling constants, 
  we introduce 
 $ g_{\phi}$, $g_{\thetatilde}$ and $g_{\phitilde}$ 
 given by 
\begin{eqnarray}   \label{eqn:expand}
K_\nu = 1 + g_\nu/2\pi\vf \,\,\,\,
 (\nu = \phi, \thetatilde, \phitilde) \virg
\end{eqnarray}  
 where $g_{\phi} =  \gpara$ and 
  $g_{\thetatilde} = g_{\phitilde} = 0$.
 By making use of  $G_z = g_z/2 \pi \vf$, 
  renormalization group equations  are obtained as 
\bleq
\begin{eqnarray}
\dl G_{\phi} (l)      
&=& - \frac{1}{2}
\left(  
    G_e^2 (l) \, J_0 (y(l))
  + G_f^2 (l)
  + G_g^2 (l)
  + G_h^2 (l)
\right)       \label{K_phi} 
 \virg    \\
\dl G_{\thetatilde} (l) 
&=& \frac{1}{2} 
\left(  
  - G_a^2 (l) \, J_0 (y(l))
  - G_b^2 (l) \, J_0 (y(l))
  + G_c^2 (l) 
  + G_d^2 (l) 
  -  g_e^2 (l) \, J_0 (y(l))
  + g_f^2 (l)
\right)\virg     \label{K_thetatilde} \\
\dl G_{\phitilde} (l)
&=& \frac{1}{2} 
\left(  
  - G_a^2 (l) \, J_0 (y(l))
  + G_b^2 (l) \, J_0 (y(l)) 
  - G_c^2 (l)    
  + G_d^2 (l) 
  - G_g^2 (l)  
  + G_h^2 (l)
\right)\virg     \label{K_phitilde}\\ 
\dl G_a (l) 
&=& 
\left(  - G_{\widetilde{\theta}}(l) - G_{\widetilde{\phi}}(l) \right)
G_a (l)
- G_e (l) \, G_g (l)        \label{g_a} 
 \virg   \\
\dl G_b (l)
&=& 
\left( - G_{\widetilde{\theta}} (l) 
        + G_{\phitilde} (l) \right)
G_b (l)
- G_e (l) \, G_h (l)        \label{g_b} 
 \virg   \\
\dl G_c (l) 
&=&
\left( + G_{\widetilde{\theta}} (l) 
        - G_{\widetilde{\phi}} (l)\right)
G_c (l)
- G_f (l) \, G_g (l)        \label{g_c} 
 \virg   \\
\dl G_d (l) 
&=&
\left( + G_{\widetilde{\theta}} (l) 
        + G_{\widetilde{\phi}} (l) \right)
G_d (l)
- G_f (l) \, G_h (l)        \label{g_d} 
 \virg   \\
\dl G_e (l)
&=&
\left(  - G_{\phi} (l) - G_{\widetilde{\theta}} (l) \right)
G_e (l)
-  G_a (l) \, G_g (l) - G_b (l) \, G_h (l)         \label{g_e} 
 \virg   \\
\dl G_f (l)
&=&
\left(  - G_{\phi} (l) + G_{\widetilde{\theta}} (l) \right)
G_f (l)
- G_c (l) \, G_g (l) - G_d (l) \, G_h (l)         \label{g_f} 
 \virg   \\
\dl G_g (l)
&=&
\left(  - G_{\phi} (l) - G_{\widetilde{\phi}} (l) \right)
G_g (l)
- G_a (l) \, G_e (l) \, J_0 (y(l)) - G_c (l) \, G_f (l)     \label{g_g} 
 \virg   \\
\dl G_h (l)
&=&
\left(  - G_{\phi} (l) + G_{\widetilde{\phi}} (l) \right)
G_h (l)
- G_b (l) \, G_e (l) \, J_0 (y(l)) - G_d (l) \, G_f (l)      \label{g_h}
 \virg   \\
               \label{g0_l} 
\dl \tt (l)
&=& 
\tt(l) 
       - \frac{1}{8} \left( G_a^2(l) + G_b^2(l) 
                 + G_e^2(l) \right) \, J_1 (y(l))  
                                    \virg 
\end{eqnarray}
\eleq
 where the detail derivation is shown in Appendix A. 
In the above equations, 
 $\tt (l) = t(l)/(\vf\al^{-1})$, $y(l) = 4\tt(l)$ and
 $J_n(y(l))$ $(n=0,1)$ is the $n$-th order Bessel function.
 Initial conditions are given by 
$
G_z (0) = g_z/2\pi\vf
$
and 
$
\tt (0) = \tn
$.
 These equations  retain the SU(2) symmetry 
  for $\gpara = \gperp = \g2$. 
 In Eqs. (\ref{g_a})-(\ref{g_h}),  
  the  bilinear term with respect to  $G_z(l)$ ($z=a \sim h$)
  does exist   for  two-coupled chains with backward scattering. 
 Such a term,  which has been  obtained by improving the previous 
  calculation \cite{Yoshioka_PRB}, 
  is  crucial  to reproduce 
  the one-dimensional renormalization  equation \cite{Solyom}
  in the limit of small $t$. 
 We examine    order parameters  for the possible states 
 in case of repulsive interaction, which are   given by 
\begin{eqnarray}
 O_{{\rm LSDW}_{\para,{\rm out}}}
 &=&         \sum_{\sigma} \sigma \,
       \{  \psi_{+,\sigma,1}^\dagger \, \psi_{-,\sigma,1} 
       - 
         \psi_{+,\sigma,2}^\dagger \, \psi_{-,\sigma,2} \}
\nonumber \\
&=& \sum_{\sigma,\mu} \sigma \, 
         \psi_{+,\sigma,\mu}^\dagger \, \psi_{-,\sigma,-\mu}
 \label{O_LSDW} 
 \virg \\ 
  O_{{\rm SS}_{\perp,{\rm in}}}
 &=&         \sum_{\sigma}   \sigma
   \{      \psi_{+,\sigma,1} \, \psi_{-,-\sigma,2} 
       + 
         \psi_{+,\sigma,2} \, \psi_{-,-\sigma,1}\}
\nonumber \\
&=& \sum_{\sigma,\mu} \sigma \mu \,
         \psi_{+,\sigma,\mu} \, \psi_{-,-\sigma,\mu}
 \label{O_SS} 
  \virg \\
 O_{{\rm TSDW}_{\para,{\rm out}}}
 &=&         \sum_{\sigma} 
       \{  \psi_{+,\sigma,1}^\dagger \, \psi_{-,-\sigma,1} 
       - 
         \psi_{+,\sigma,2}^\dagger \, \psi_{-,-\sigma,2} \}
\nonumber \\
&=& \sum_{\sigma,\mu}
    \psi_{+,\sigma,\mu}^\dagger \, \psi_{-,-\sigma,-\mu}
 \label{O_TSDW} 
  \virg \\
 O_{{\rm CDW}_{\perp,{\rm out}}}
 &=&         \sum_{\sigma}
       \{  \psi_{+,\sigma,1}^\dagger \, \psi_{-,\sigma,2} 
       - 
         \psi_{+,\sigma,2}^\dagger \, \psi_{-,\sigma,1} \}
\nonumber \\
&=& \sum_{\sigma,\mu} \mu \,
    \psi_{+,\sigma,\mu}^\dagger \, \psi_{-,\sigma,-\mu} 
 \label{O_CDW} 
  \virg \\ 
 O_{{\rm TS}_{\para,{\rm in}}}
 &=&         \sum_{\sigma} 
   \{      \psi_{+,\sigma,1} \, \psi_{-,-\sigma,1} 
       + 
         \psi_{+,\sigma,2} \, \psi_{-,-\sigma,2}\}
\nonumber \\
&=& \sum_{\sigma,\mu}
    \psi_{+,\sigma,\mu} \, \psi_{-,-\sigma,\mu}
  \virg 
 \label{O_TS} 
\end{eqnarray}
where   
$\psi_{p,\sigma,i} (x) = (1/\sqrt{L} ) 
 \sum_k \e^{ikx} a_{k,p,\sigma,i}$. 
 In Eqs. (\ref{O_LSDW})-(\ref{O_TS}), 
 ${\rm LSDW}_{\para,{\rm out}}$  (${\rm TSDW}_{\para,{\rm out}}$) 
 and ${\rm CDW}_{\para,{\rm out}}$ 
 denote  longitudinal (transverse) SDW 
 with intrachain and out-of-phase pairing 
 and 
 charge density wave (CDW) 
 with intrachain and out-of-phase pairing 
 while 
  ${\rm SS}_{\perp,{\rm in}}$ (${\rm TS}_{\perp,{\rm in}}$)  
 represents 
the singlet (triplet) SC state with interchain and in-phase pairing. 
  Since there is   a  symmetry with respect to $\gpara= 2g_2$ 
  in Eq. (\ref{phase_Hamiltonian}),
 the case with $ 2 g_2 > \gpara > 0$ is investigated explicitly 
  where the relevant order parameters in terms of phase variables are   reexpressed as
\bleq
\begin{eqnarray}              \label{order_LSDW}
          O_{{\rm LSDW}_{\para,{\rm out}}}(x)  
&\to&
        \e^{-i2\kf x}
         \displaystyle{\sum_{\sigma}}   
         \exp \left[ -i
           \bigl\{ \theta_+ + \sigma \phi_+
           \bigr\}/\sqrt{2}
              \right] \,
         \cos \left[
         \left\{ \thetatilde_- + \sigma \phitilde_- \right\}/\sqrt{2}
         \right]   
       \virg \\
\label{order_TSDW}
          O_{{\rm TSDW}_{\para,{\rm out}}}(x)  
&\to&
      \e^{-i2\kf x}
         \displaystyle{\sum_{\sigma}} \sigma   
         \exp \left[ -i
           \bigl\{ \theta_+ + \sigma \phi_-
           \bigr\}/\sqrt{2}
              \right] \,
         \sin \left[
         \left\{ \thetatilde_- + \sigma \phitilde_+ \right\}/\sqrt{2}
         \right]   
       \virg \\
                   \label{order_SS}
       O_{{\rm SS}_{\perp,{\rm in}}}(x)
&\to&
      \sum_{\sigma}  \sigma \,
          \exp \left[ i
               \bigl\{ \theta_- + \sigma \phi_+  \bigr\} /\sqrt{2}
               \right] \,
          \sin \left[
               \left\{ \thetatilde_- + \sigma\phitilde_+ \right\}/
         \sqrt{2}
          \right] 
       \point 
\end{eqnarray}
\bcols
We study only  2$\kf$-SDW   and  SC states and do not consider 
 4$\kf$-CDW  found for $K_{\theta}<1/2$,  
 of which denotes the range  beyond that of the conventional Hubbard model \cite{Schulz}.

\section{SS state vs. SDW state}
 We examine the state in the limit of low energy by calculating 
  Eqs. (\ref{K_phi})-(\ref{g0_l}) where the corresponding energy is 
  given by $\ef \e^{-l}$.
The numerical calculation is performed with the fixed $t/\ef = 0.1$ 
 or $0.01$ 
  and several choices of coupling constants in the region of 
  $2 g_2 > \gpara > 0$.
 The quantity $\ef$($=\vf\al^{-1}$) denotes a Fermi energy
  and the coupling constant, 
 which has the magnitude of  the order of the bandwidth, 
 is given by $\pi\vf$.
 Equations (\ref{K_phi}), (\ref{K_thetatilde}) and (\ref{K_phitilde})
 show that 
  coupling constants corresponding to the total spin,
  the transverse charge and the transverse spin are renormalized 
  leading to the excitation gap for the relevant case 
  while that of the total charge is unrenormalized. 
 We find that, among Eqs. (\ref{order_LSDW}),
  (\ref{order_TSDW}) and (\ref{order_SS}),  
  LSDW (TSDW) state becomes dominant for the relevant $\phi_+$, $\thetatilde_-$
  and $\phitilde_-$ ($\phi_-$, $\thetatilde_-$ and $\phitilde_+$) and 
  SC state becomes dominant for 
the relevant $\phi_+$, $\thetatilde_-$ and   $\phitilde_+$.

\begin{figure}[top]
\vspace*{-0.cm}\hspace{.25cm}
\epsfxsize=2.5in\epsfbox{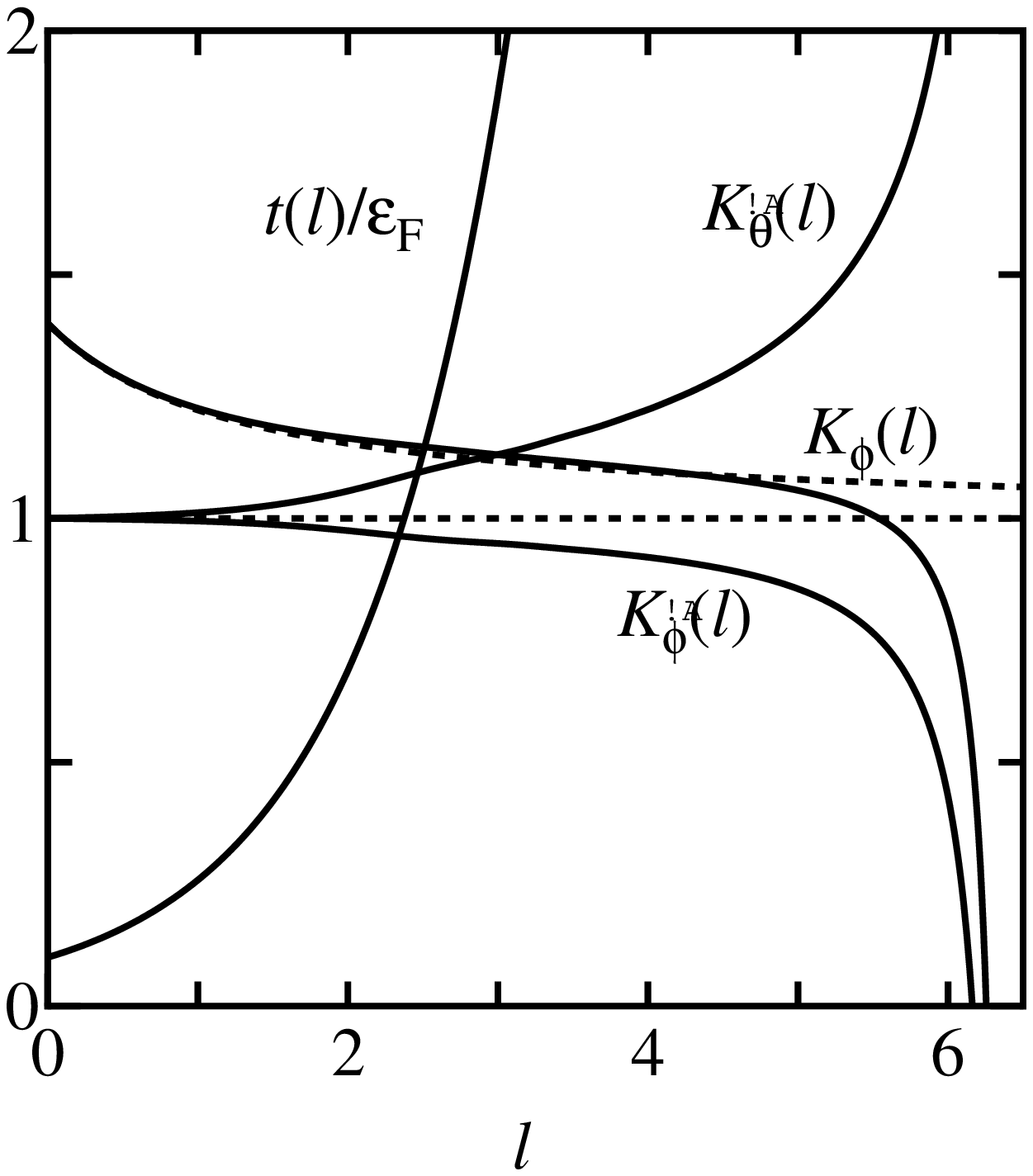}

{\noindent FIG. 1:
 The $l$-dependence of  
  $K_{\thetatilde}(l)$,  $K_{\phitilde}(l)$, $K_{\phi}(l)$ 
  and $t(l)/\ef$ for $t/\ef=0.1$ and 
   $\gpara/2\pi\vf = \gperp/2\pi\vf = g_2/2\pi\vf = 0.4$,
  where $K_{\nu}(l) \equiv 1 + G_\nu (l)$ 
  ($\nu = \thetatilde, \phitilde, \phi$).
  The dotted curves denote $K_{\nu}(l)$ of  one-dimensional case 
 given by $t=0$. 
}
\end{figure}
\noindent

 First the case with spin-isotropic interaction is shown 
 in Fig. 1 where $\gpara/2\pi\vf = \gperp/2\pi\vf = g_2/2\pi\vf = 0.4$.
 In this case,  response function of LSDW is the same as that of 
  TSDW for $\gpara = \gperp$ due to the SU(2) symmetry 
 for the spin rotation.
 The quantity $t(l)$ is always relevant i.e. increases steeply 
  for the present choice of parameters.  
 Renormalization of coupling constants shows that $G_c(l)$ and $G_f(l)$
  and $G_g(l)$ become relevant while $G_a(l)$, $G_b(l)$ and $G_e(l)$ 
  become irrelevant due to the spatially oscillating term in 
  Eq. (\ref{phase_Hamiltonian}) and both 
  $|G_d(l)/G_c(l)|$ and $|G_h(l)/G_g(l)|$ reduce to zero.
Since $K_\phi (l)$ and $K_{\phitilde}(l)$ decrease 
  and $K_{\thetatilde} (l)$ increases,
  it turns out that the fluctuations for $\phi_+$, $\phitilde_+$ 
 and $\thetatilde_-$  become relevant.
The resultant excitation gap leads to a locking of phases at 
  $\sqrt{2}\phi_+ =0$, $\sqrt{2}\thetatilde_- =0$ and 
  $\sqrt{2}\phitilde_+ =\pi$ (or $\sqrt{2}\phi_+ =\pi$, 
  $\sqrt{2}\thetatilde_- =\pi$ and $\sqrt{2}\phitilde_+ =0$ )
  because the signs of the renormalized coupling constants 
 for large $l$ are given by 
  $G_c(l)>0$, $G_f(l)<0$ and $G_g(l)>0$. 
Thus the dominant state in the limit of low energy is given by SC state
  with the singlet pairing of interchain and in-phase. 
 Such a result of SC state is consistent with the previous one
\cite{Fabrizio,Balents,Schulz}.
 For parameters used in   Fig. 1, 
 SS (SDW) state  becomes dominant compared to SDW (SS) state 
 when  the energy is  smaller (larger) than 
 a magnitude of $\ef \exp[-6.4]$. 
 For the comparison, the result for a single chain (i.e., $t=0$) 
 is shown by the dotted curve 
  where $K_{\thetatilde}(l)=K_{\phitilde}(l)=1$ and $K_\phi(l)$ 
  decreases to 1.
The difference between the solid curve and the dotted curve 
 is attributable to  the interchain hopping and appears 
for  $l \gsim 1$,   which corresponds to $4t(l)/\ef \gsim 1$.

\begin{figure}[t]
\vspace*{-0.cm}\hspace{.05cm}
\epsfxsize=2.8in\epsfbox{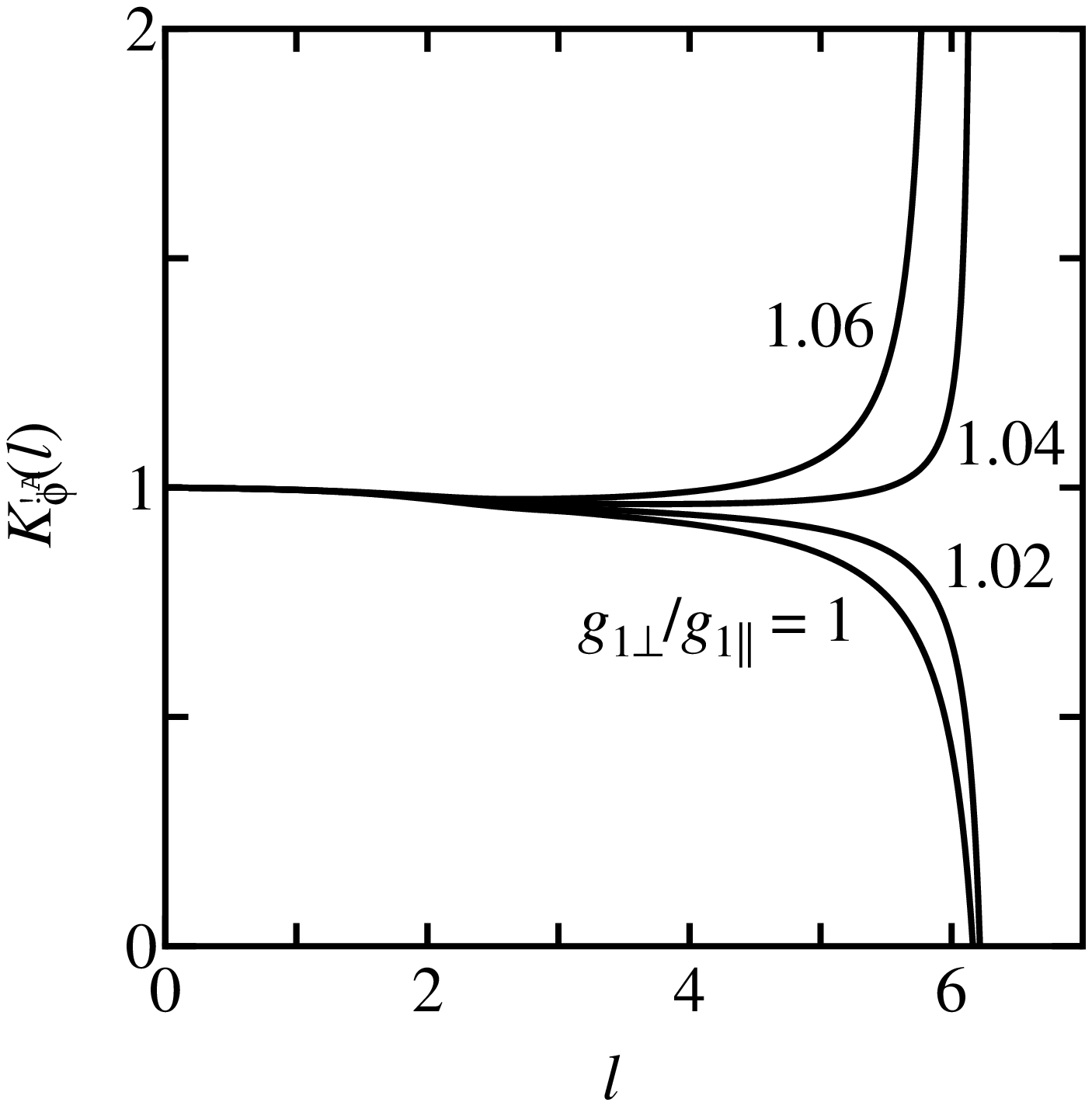}

{\noindent FIG. 2:
 The $l$-dependence of $K_{\phitilde}(l)$ with the fixed 
  $\gperp/\gpara = 1$, $1.02$, $1.04$ and $1.06$ 
  where $t/\ef = 0.1$ and $\gpara/2\pi\vf = g_2/2\pi\vf = 0.4$.
}
\end{figure}
\noindent

Next  the state for the spin-anisotropic interaction is examined 
  by varying the magnitude of the backward scattering with 
  opposite spins, $\gperp$. 
 The case of $\gperp / \gpara >1$ is calculated  
 where response function for LSDW becomes larger than that for TSDW. 
The case of $\gperp / \gpara <1$ is discussed in section 4. 
In Fig. 2, the $l$-dependence of $K_{\phitilde}(l)$ is shown by
  choosing $\gperp/\gpara= 1$, $1.02$, $1.04$ and $1.06$ 
  where $\gpara/2\pi\vf = g_2/2\pi\vf = 0.4$ and $t/\ef = 0.1$.
There are two kinds of fixed points depending on 
the magnitude of $\gperp/\gpara$. 
 The quantity $K_{\phitilde}(l)$ with large $l$ decreases for 
  $\gperp/\gpara <1.035$ (case (I)) 
  but increases  otherwise (case (II))
  although $K_{\phitilde}(l)$ with small $l$ always decreases.
In case (I), the behavior of renormalization 
 is  similar to that of 
  the spin-isotropic case and then SC state becomes a dominant state.
 In  case (II), the rapid increase of $K_{\phitilde} (l)$ shows 
  the relevance $\phitilde_-$ instead of $\phitilde_+$ and 
 then  leads to the reduction of  
  $|G_c(l)/G_d(l)|$ and $|G_g(l)/G_h(l)|$ which is opposite to 
 the case (I).
From Eq. (\ref{order_LSDW}),  the dominant state 
 in the limit of low energy for the case (II) 
 is given by LSDW with intrachain and out-of-phase pairing 
  since the signs of relevant couplings are given by $G_d(l)<0$, 
  $G_f(l)<0$ and $G_h(l)<0$ and then phases are locked by 
  $\sqrt{2}\phitilde_- = 0$, $\sqrt{2}\phi_+=0$ and  
  $\sqrt{2}\thetatilde_- = 0$ (or $\sqrt{2}\phitilde_- = \pi$, 
  $\sqrt{2}\phi_+=\pi$ and $\sqrt{2}\thetatilde_- = \pi$). 
 In the one-dimensional case, the dominant state 
with $\gperp > \gpara$ is given by  LSDW
\cite{Suzumura_JPSJ84,Giamarchi_JPF} due to  the excitation gap 
  induced in the spin fluctuation \cite{Emery}. 
 For two-coupled chains with $\gperp > \gpara$, the difference 
  between response function of LSDW and that of TSDW is enhanced 
 for $l\gsim |\ln(4t^*/\ef)|$,
 where  $t$ is renormalized as $t^*$
  due to one-dimensional fluctuation
\cite{Bourbonnais_MCLC85,Tsuchiizu}. 
  In the case (II),  the dominant state is given by LSDW 
 for all the  energies  while a crossover from LSDW to SC state 
  with decreasing energy is obtained in the case (I).

\begin{figure}[t]
\vspace*{-0.cm}\hspace{.05cm}
\epsfxsize=2.8in\epsfbox{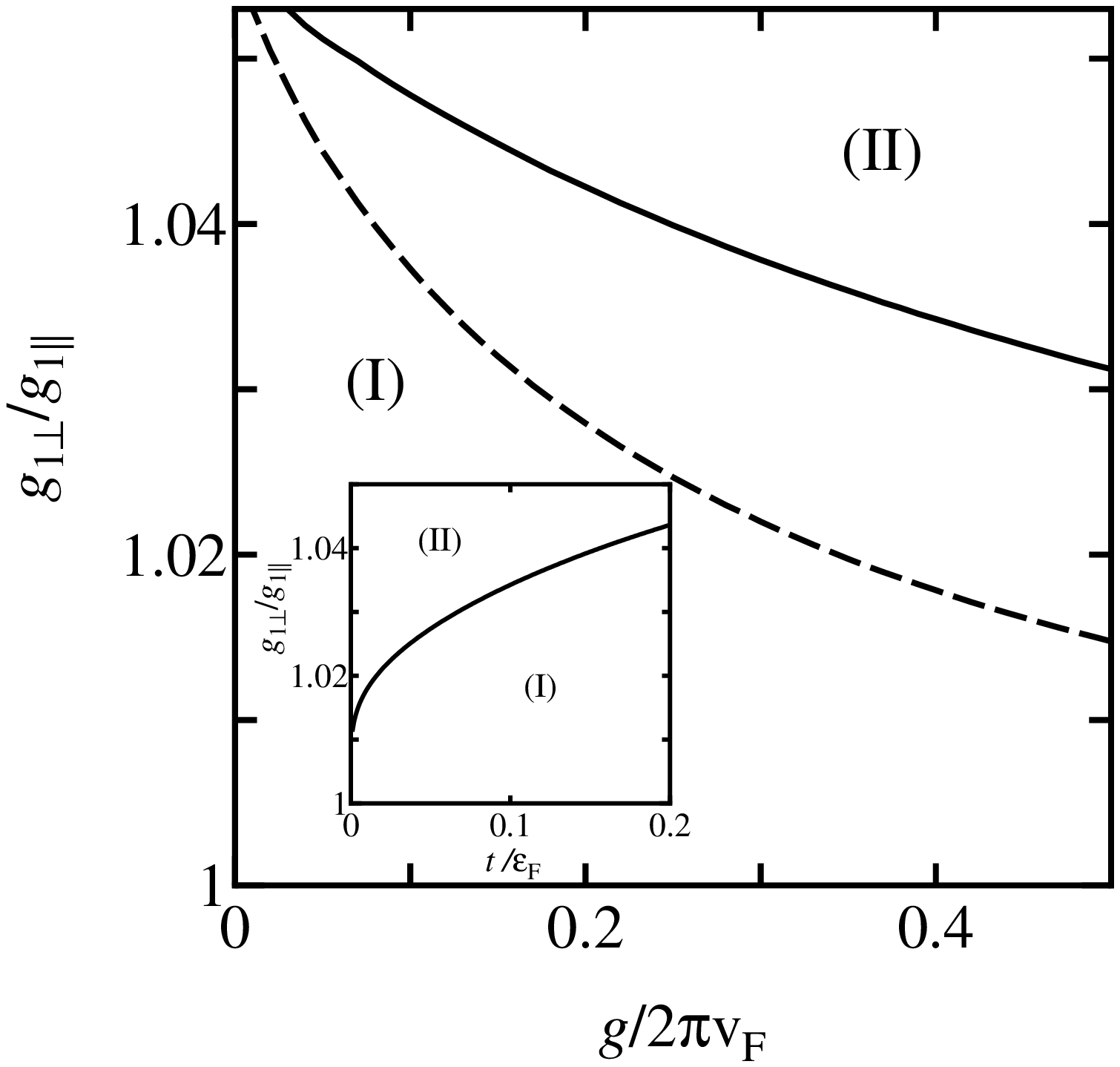}

{\noindent FIG. 3:
 Phase diagram  on the plane of  
  $g$ and $\gperp/\gpara$ where 
 where  $g \equiv g_2 = \gpara$.
 The regions (I) and (II) correspond to  SC state  SDW state 
 respectively  and  solid (dotted) curve  denotes  the boundary 
 for  $t/\ef = 0.1$ ( $t/\ef = 0.01$). 
In the inset,
  phase diagram on the plane of $t$ and $\gperp/\gpara$ 
 with the fixed  $g/2\pi\vf = 0.4$ is shown.
}
\end{figure}
\noindent

In Fig. 3, the ground state is shown on the plane of 
 $g$ and $\gperp/\gpara$ 
 with the fixed $t/\ef = 0.1$ (solid curve)
  and $0.01$ (dashed curve) where $g$($=\gpara=g_2$).
 The regions (I) and  (II) denote
  SS$_{\perp,{\rm in}}$ and LSDW$_{\para,{\rm out}}$, respectively.
 For $\gperp/\gpara = 1$, we always obtain the SC state. 
  For  $\gperp/\gpara > 1$, 
 there is a critical value of  $g$  at which 
  the SC state moves  to the LSDW state with increasing $g$.  
 Such a critical value decreases with increasing  
 $\gperp/\gpara$ and/or decreasing $t/\ef$. 
The inset shows a phase diagram of SC state (I) and LSDW state (II)
  on the plane of $t$ and $\gperp/\gpara$  with the fixed 
 $g/2\pi\vf = 0.4$ where the SC (LSDW) state is obtained 
  with increasing interchain hopping (anisotropy of interaction).
For SC state, there is an energy gain by the $g_c$-term which 
  competes with the $g_a$-term and then becomes relevant for energy 
  smaller than $t$.
On the other hand, the energy gain for LSDW comes from the $g_h$-term,
  which is smaller but cooperates  with the $g_d$-term 
 for all the energies.

\begin{figure}[t]
\vspace*{-0.cm}\hspace{.05cm}
\epsfxsize=2.8in\epsfbox{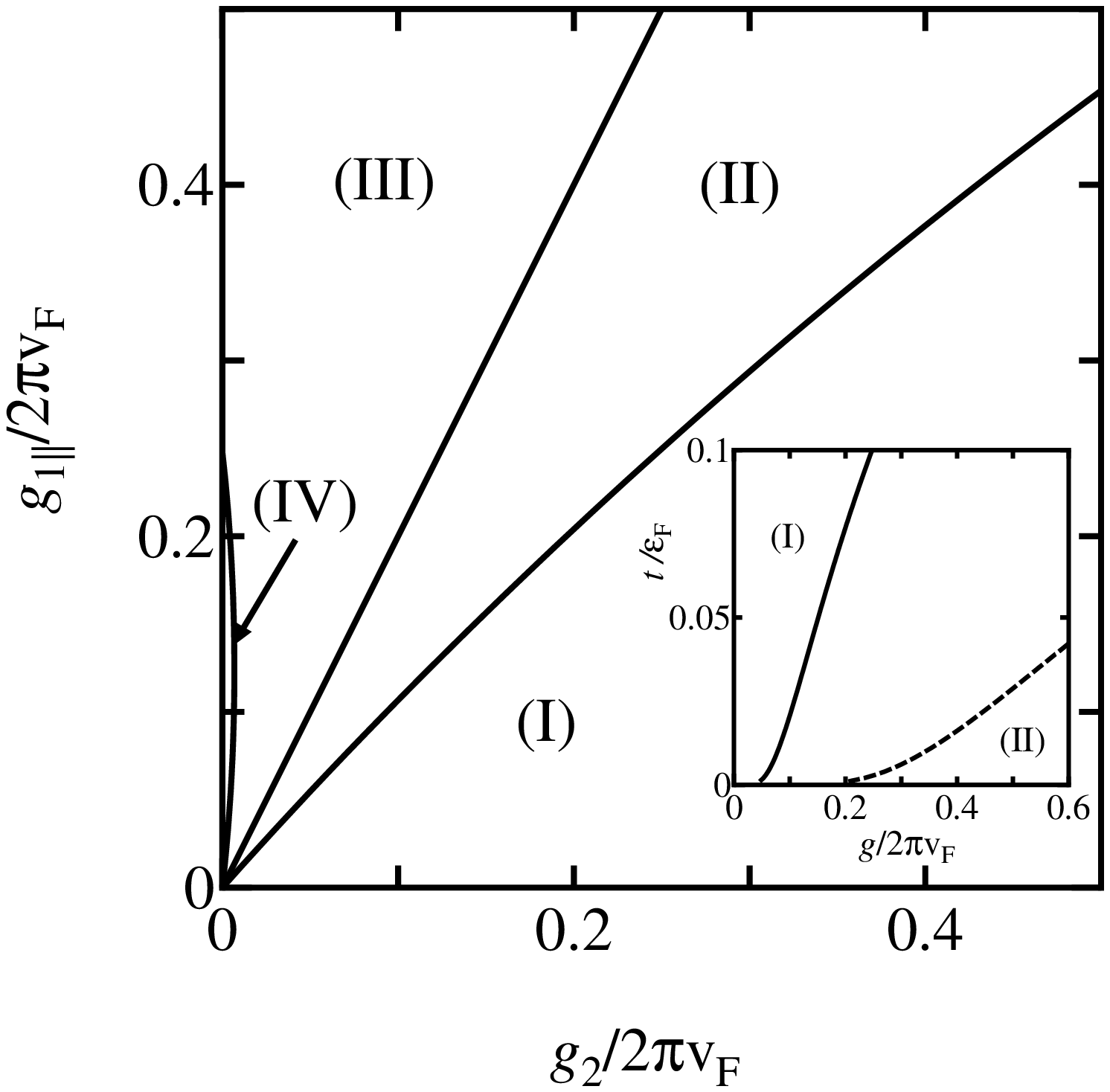}

{\noindent FIG. 4:
  Phase diagram for two-coupled chains 
  on the plane of $g_2$ and $\gpara$  
  where $t/\ef=0.1$ and $\gperp/ \gpara = 1.04$. 
 Regions (I), (II), (III) and (IV) 
denote SS,  SDW, TS and CDW states which  are given by 
Eqs. (\ref{O_SS}), (\ref{O_LSDW}), (\ref{O_TS}) and (\ref{O_CDW}) 
  respectively.  
In the inset, the phase diagram of SS (I) and SDW (II) state 
 on the plane of $g$ and $t$ in the case of 
 $g_2 = \gpara \equiv g$ is shown 
 where the solid curve and dashed curve denote the boundaries for
 $\gperp/ g = 1.04$ and $1.02$ respectively.
}
\end{figure}
\noindent

 In Fig. 4, the phase diagram is shown on the plane of $g_2$ 
and $\gpara$
  with the fixed $\gperp/\gpara = 1.04$ and $t/\ef = 0.1$.
The regions (I), (II), (III) and (IV) correspond to states 
 for SS$_{\perp,{\rm in}}$, LSDW$_{\para,{\rm out}}$,
  TS$_{\para,{\rm in}}$ and CDW$_{\perp,{\rm out}}$, 
 which are given by 
Eqs. (\ref{O_SS}), (\ref{O_LSDW}), (\ref{O_TS}) and (\ref{O_CDW}) 
  respectively.
There is a symmetry given by 
  $g_2 \to \gpara - g_2$ with the fixed $\gpara$ 
 for the magnitude of response functions  
  where response function of LSDW becomes equal to that of TS
  at $\gpara = 2 g_2$.
In the inset, the phase diagram of SS$_{\perp,{\rm in}}$ (I) and
  LSDW$_{\para,{\rm out}}$ (II) on the plane of $g$ - $t$ is shown 
  for  $\gperp / \gpara = 1.04$ (solid curve) and 
   $\gperp / \gpara = 1.02$ (dashed curve).
Since these curves are well reproduced by the formula,
  $t/\ef = \exp[ - C_1 - C_2 \cdot 2\pi\vf /g]$,  with 
  $C_1 \simeq 1.3 (1.2)$ and   $C_2 \simeq 2.6 (1.2)$ 
for $\gperp/\gpara = 1.04 (1.02)$, 
  it is found that the boundary is determined by the competition
  between the interchain hopping and the spin gap induced by the 
  spin-anisotropy interaction.

\section{Discussion} 
 By applying the renormalization group method within the second order, 
 we have investigated the dominant state 
  in the limit of low energy 
 for two-coupled chains with the spin-anisotropic backward scattering. 
 The   phase diagram where  SDW and SS states
   are calculated as the function of 
   $\gpara$, $g_2$ and $\tilde{t}$ shows 
    a noticeable   interplay 
  of   the interchain hopping and the intrachain interaction. 

In the previous section, we have examined the case $\gperp > \gpara$, 
which leads to a competition between SC state and SDW state. 
 For $\gperp < \gpara$, we have found that  results are always 
  similar to those of Fig. 1 and then 
 the dominant state is given by SC state. 
 The difference between the case of  $\gperp > \gpara$ and 
 that of $\gperp < \gpara$ 
  originates in the fact that
  the renormalization for $\gperp > \gpara$ ($\gperp < \gpara$)
  leads to a strong (weak coupling) coupling  regime
  in a one-dimensional model \cite{Emery}. 
   The case of $\gperp > \gpara$, which leads to  LSDW state, 
   denotes interactions which   enhance 
    the anisotropy along the direction parallel to 
  the quantized axis.
When the anisotropy is large for the perpendicular direction,
  one can expect TSDW in a similar way by adding 
     interactions  with   the   perpendicular anisotropy,
  which have  been treated 
  in the one-dimensional model \cite{Giamarchi_JPF}.

  We compare the present result with that 
  of Fabrizio \cite{Fabrizio} 
  who obtained the SC state 
  for two-coupled chains of 
  the Hubbard model by use of a diagrammatic method.
 The relation between  our coupling constants and those of him 
 is   given by 
  $G_\theta = g^\para + {g_f}^\para - g^{(2)} - {g_f}^{(2)}$,
  $G_\phi   = g^\para + {g_f}^\para + g^{(2)} + {g_f}^{(2)}$,
  $G_{\thetatilde} = g^\para - {g_f}^\para - g^{(2)} + {g_f}^{(2)}$,
  $G_{\phitilde}   = g^\para - {g_f}^\para + g^{(2)} - {g_f}^{(2)}$,
  $G_a = -2 {g_b}^\para$, $G_b = -2 {g_b}^{(2)}$, 
$G_c = 2 {g_t}^{(2)}$,
  $G_d = 2{g_t}^\para$, $G_e = 2{g_b}^{(1)}$, $G_f = -2{g_t}^{(1)}$,
  $G_g = -2g^{(1)}$ and $G_h = -2{g_f}^{(1)}$. 
He has derived eight  equations by taking account of  
 the  SU(2) symmetry in the spin space, 
 with the condition, ${g_i}^\para = {g_i}^{(1)} - {g_i}^{(2)}$.
 We have obtained eleven renormalization group equations,
  i.e. Eqs. (\ref{K_phi})-(\ref{g_h}), for coupling 
  constants 
  which become equal to his results within the second order
  under the assumption of the symmetry. 
Our renormalization group equations has been derived  
   to cover  the space where  
  the spin-anisotropic interaction can be treated.

Finally, we comment on the anisotropy of SDW state  
  of organic conductors 
  where the experimental evidence has been shown 
  by spin susceptibility \cite{Gruner} 
  and the origin has been discussed in terms of 
  spin-orbit coupling and dipole-dipole interaction 
  \cite{Giamarchi_JPF}.
In the present paper, we have shown in Figs. 3 and 4 that a transition
  from SDW state to SC state occurs in the presence of spin-anisotropic
  backward scattering by the increase of the interchain hopping 
  and/or the decrease of the renormalized coupling constant.
Although such a transition might  be expected  under pressure 
  resulting in the increase of band width and then $\vf$,
  the critical value for $(\gperp - \gpara)/\gpara$,
  which is evaluated by the moderate choices 
    of parameters  (i.e., $g \simeq \pi \vf$ and $t/\ef =0.1$), 
  is much larger than that found in the experiment \cite{Gruner}.
It is rather plausible to consider that 
 the transition from SDW state 
  to SC state in organic conductors \cite{Jerome} comes from 
  the suppression of the nesting condition.
However the results of the present calculation indicate 
  a reasonable magnitude of  spin-anisotropic interaction 
   for the competition between  SDW state and SC state,
  which may be found in quasi-one-dimensional systems.

\vspace{.5cm}
{\bf Acknowledgment}  \\
 The authors thank to H. Yoshioka for useful discussions. 
 This work was partially supported by  a Grant-in-Aid 
 for Scientific  Research  from the Ministry of Education, 
Science, Sports and Culture, (No.09640429) Japan. 

\bleq
\appendix
\section{Derivation of renormalization group equations}
\bcols
By treating the nonlinear terms 
  in Eq. (\ref{phase_Hamiltonian})
  as the perturbation, the response function for 
  $\phi_+$ field is calculated
  up to the third order as
\bleq
\begin{eqnarray}
&&
\lan T_\tau \e^{ (i/\sqrt{2})\phi_+ (x_1,\tau_1)} 
               \e^{-(i/\sqrt{2})\phi_+ (x_2,\tau_2)} \ran \nonumber \\
&=&  \e^{-(K_\phi/2)U (r_1-r_2)}    \nonumber \\
&&  + \frac{1}{2!} \frac{1}{(4\pi)^2} 2 \sum_{\epsilon=\pm 1} 
     \int \frac{\d ^2 r_3}{\alpha^2}\frac{\d ^2 r_4}{\alpha^2}
     \e^{-(K_\phi/2) U (r_1-r_2)} 
     \e^{-2K_\phi    U (r_3-r_4)} \nonumber \\
&& \hspace{0.5cm} \times 
   \left(   \e^{ 
               \epsilon K_\phi \left\{
    U(r_1-r_3)-U(r_1-r_4)-U(r_2-r_3)+U(r_2-r_4)
               \right\}
           } -1 \right)
      \nonumber \\ 
&& \hspace{.5cm} \times \left\{
            G_e^2 \e^{-2K_{\thetatilde}U(r_3-r_4)} 
                       \cos 2q_0 (x_3-x_4)
          + G_f^2 \e^{-(2/K_{\thetatilde})U(r_3-r_4)}
          + G_g^2 \e^{-2K_{\phitilde}U(r_3-r_4)}
          + G_h^2 \e^{-(2/K_{\phitilde})U(r_3-r_4)}
                       \right\}  \nonumber \\
&& - \frac{1}{3!} \frac{1}{(4\pi)^3} 4 \sum_{\epsilon = \pm 1} 
     \int \frac{\d ^2 r_3}{\alpha^2} 
          \frac{\d ^2 r_4}{\alpha^2}
          \frac{\d ^2 r_5}{\alpha^2}
     \e^{-(K_\phi/2)U(r_1-r_2)} \e^{-2K_\phi U(r_3-r_4)} 
          \nonumber \\
&& \hspace{0.5cm} \times 
     \left(
       \exp \left[
               \epsilon K_\phi \biggl\{
       U(r_1-r_3)-U(r_1-r_4)-U(r_2-r_3)+U(r_2-r_4)
               \biggr\}
            \right] -1
     \right) \nonumber \\
&& \hspace{.5cm} \times \left\{
          6 G_a G_e G_g \e^{-2K_{\thetatilde}U(r_3-r_5)} 
                        \e^{-2K_{\phitilde}U(r_4-r_5)} 
                        \cos 2q_0 (x_3-x_5)
                       \right.  \nonumber \\ 
&& \hspace{2.5cm}
        + 6 G_b G_e G_h \e^{-2K_{\thetatilde}U(r_3-r_5)}
                        \e^{-(2/K_{\phitilde})U(r_4-r_5)}
                         \cos 2q_0 (x_3-x_5)
                         \nonumber \\
&& \hspace{1cm} \left.
        + 6 G_c G_f G_g \e^{-(2/K_{\thetatilde})U(r_3-r_5)}
                        \e^{-2K_{\phitilde}U(r_4-r_5)}
        + 6 G_d G_f G_h \e^{-(2/K_{\thetatilde})U(r_3-r_5)}
                        \e^{-(2/K_{\phitilde})U(r_4-r_5)}
                       \right\}   + \ldots
\virg        \label{A1}
\end{eqnarray}
\eleq
  where 
  $U (r) = \ln \left[\sqrt{x^2 + \vf^2 \tau^2}/\alpha\right]$, 
   $\d^2r = \vf \, \d x \, \d \tau$, $x = x_1 - x_2$ and 
  $\tau = \tau_1 - \tau_2$.
The quantity $v_\phi$ is replaced by $\vf$ and $q_0=2t/\vf$.
In order to obtain scaling equations 
  of the coupling constants up to the second order,
  we need the response functions expanded up to the third order 
  of the nonlinear terms,
  while these third terms are absent in 
  one-dimensional case \cite{Giamarchi_JPF}.
  These results originate in the nonlinear terms 
  including both $\phi_+$ and $\thetatilde_{\pm}$ 
  (or $\phi_+$ and $\phitilde_{\pm}$)
  in Eq. (\ref{phase_Hamiltonian}).
By putting $r_5=r_4+r$ and $r_5=r_3+r$, and expanding near $r=0$,
  we find
\bleq
\begin{eqnarray}
{\rm Eq. \,\, (\ref{A1})}
&=&  \e^{-(K_\phi/2)U(r_1-r_2)}   
  + \frac{1}{(4\pi)^2} \sum_{\epsilon} 
     \int \frac{\d ^2 r_3}{\alpha^2} 
          \frac{\d ^2 r_4}{\alpha^2}
     \e^{-(K_\phi/2)U(r_1-r_2)} \e^{-2K_\phi U(r_3-r_4)} 
          \nonumber \\
&& \hspace{0.5cm} \times 
     \left(
       \exp \left[
               \epsilon K_\phi \biggl\{
        U(r_1-r_3)-U(r_1-r_4)-U(r_2-r_3)+U(r_2-r_4)
               \biggr\}
            \right] -1
     \right) \nonumber \\
&& \hspace{0.5cm} \times \left\{
            {G_e^{\rm eff}}^2 
                  \e^{-2K_{\thetatilde}U(r_3-r_4)} 
                       \cos 2q_0 (x_3-x_4)
          + {G_f^{\rm eff}}^2  
                       \e^{-(2/K_{\thetatilde})U(r_3-r_4)}
                       \right.  \nonumber \\
&& \hspace{4.5cm}        \left.
          + {G_g^{\rm eff}}^2 \e^{-2K_{\phitilde}U(r_3-r_4)}
          + {G_h^{\rm eff}}^2 \e^{-(2/K_{\phitilde})U(r_3-r_4)}
                       \right\}  \label{A3} \point 
\end{eqnarray}
The quantities $G_z^{\rm eff}$ are given by
\begin{eqnarray}
{G_e^{\rm eff}}^2 &=& G_e^2 
    -2 G_a G_e G_g \int \frac{\d r}{\alpha} 
           \left( \frac{r}{\alpha} \right)^{1-2K_{\phitilde}}
    -2 G_b G_e G_h \int \frac{\d r}{\alpha} 
           \left( \frac{r}{\alpha} \right)^{1-2/K_{\phitilde}}\virg 
\label{eqn:G_e} \\
{G_f^{\rm eff}}^2 &=& G_f^2 
    -2 G_c G_f G_g \int \frac{\d r}{\alpha} 
           \left( \frac{r}{\alpha} \right)^{1-2K_{\phitilde}}
    -2 G_d G_f G_h \int \frac{\d r}{\alpha} 
           \left( \frac{r}{\alpha} \right)^{1-2/K_{\phitilde}}\virg  
\label{eqn:G_f} \\
{G_g^{\rm eff}}^2 &=& G_g^2 
    -2 G_a G_e G_g \int \frac{\d r}{\alpha} 
           \left( \frac{r}{\alpha} \right)^{1-2K_{\thetatilde}}
           J_0(2q_0 r)
    -2 G_c G_f G_g \int \frac{\d r}{\alpha} 
           \left( \frac{r}{\alpha} \right)^{1-2/K_{\thetatilde}}\virg 
\label{eqn:G_g} \\
{G_h^{\rm eff}}^2 &=& G_h^2 
    -2 G_b G_e G_h \int \frac{\d r}{\alpha} 
           \left( \frac{r}{\alpha} \right)^{1-2K_{\thetatilde}}
           J_0(2q_0 r)
    -2 G_d G_f G_h \int \frac{\d r}{\alpha} 
           \left( \frac{r}{\alpha} \right)^{1-2/K_{\thetatilde}}
\label{eqn:G_h} \virg
\end{eqnarray}
\eleq
where $r = (x^2 + (\vf \tau)^2)^{1/2}$ and 
  the second and third terms of r.h.s. 
  in Eqs. (\ref{eqn:G_e})-(\ref{eqn:G_h}) are obtained by exponentiating
  the third order terms of Eq. (\ref{A1}).
From Eq. (\ref{A3}), the quantity $K_\phi^{\rm eff}$ 
  is derived as 
\bleq
\begin{eqnarray}
K_\phi^{\rm eff} &=& K_\phi
-\frac{1}{2} G_e^2 K_\phi^2 \int \frac{\d r}{\alpha}
                  \left( \frac{r}{\alpha} \right)
                   ^{3-2K_\phi-2K_{\thetatilde}}
                  J_0(2q_0r) 
                  U(r_1-r_2)        \nonumber \\
&-& \hspace{0cm} 
      \frac{1}{2} G_f^2 K_\phi^2 \int \frac{\d r}{\alpha}
                  \left( \frac{r}{\alpha} \right)
                   ^{3-2K_\phi-2/K_{\thetatilde}}
                  U(r_1-r_2)    
    - \frac{1}{2} G_g^2 K_\phi^2 \int \frac{\d r}{\alpha}
                  \left( \frac{r}{\alpha} \right)
                   ^{3-2K_\phi-2K_{\phitilde}}
                  U(r_1-r_2)     \nonumber \\
&-& \hspace{0cm}
      \frac{1}{2} G_h^2 K_\phi^2 \int \frac{\d r}{\alpha}
                  \left( \frac{r}{\alpha} \right)
                   ^{3-2K_\phi-2/K_{\phitilde}}
                  U(r_1-r_2)  \point   \label{eqn:app-Kphi}
\end{eqnarray}
\eleq
For  $\alpha \to \alpha' = \alpha \e^{\d l}$ \cite{Giamarchi_JPF},
these quantities are scaled as
$
K_\phi^{\rm eff}(K_\nu ',g',q_0',\alpha') 
  = K_\phi^{\rm eff}(K_\nu,g,q_0,\alpha)
$,
$
G_e^{\rm eff}(K_\nu ',G_z',q_0',\alpha') 
  = G_e^{\rm eff}(K_\nu,G_z,q_0,\alpha)$
$   \left( \alpha'/\alpha \right)^{2-K_\phi-K_{\thetatilde}}
$,
$
G_f^{\rm eff}(K_\nu ',G_z',q_0',\alpha') 
  = G_f^{\rm eff}(K_\nu,G_z,q_0,\alpha)$
$    \left( \alpha'/\alpha \right)^{2-K_\phi-1/K_{\thetatilde}}
$,
$
G_g^{\rm eff}(K_\nu ',G_z',q_0',\alpha') 
  = G_g^{\rm eff}(K_\nu,G_z,q_0,\alpha)$
$    \left( \alpha'/\alpha \right)^{2-K_\phi-K_{\phitilde}}
$ and 
$
G_h^{\rm eff}(K_\nu ',G_z',q_0',\alpha') 
  = G_h^{\rm eff}(K_\nu,G_z,q_0,\alpha)$
$    \left( \alpha'/\alpha \right)^{2-K_\phi-1/K_{\phitilde}}
$
  where $K_\nu '$, $G_z'$ and $q'_0$ denote renormalized quantities.
From this infinitesimal transform 
  and putting $K_\nu(l)^{\pm 1} = 1 \pm G_\nu(l)$,
  the renormalization equations for the coupling constants,
  Eqs. (\ref{K_phi}), (\ref{g_e}), (\ref{g_f}), (\ref{g_g}) and 
  (\ref{g_h}) are given by Eqs. (\ref{eqn:app-Kphi}),
  (\ref{eqn:G_e}), (\ref{eqn:G_f}), 
  (\ref{eqn:G_g}) and (\ref{eqn:G_h}) respectively.
In a similar way,
  renormalization group equation for 
  $K_{\thetatilde}(l)$ ($K_{\phitilde}(l)$) is calculated
  from the response function for $\thetatilde$ ($\phitilde$) field,
and Eqs. (\ref{g_a})-(\ref{g_d}) are obtained from 
  the response function for both $\thetatilde$ and $\phitilde$ fields.

The renormalization equation for $t(l)$ is calculated as follows.
The quantity, $\Delta n$, which denotes the
 actual difference of the density between two bands,
  having $k_{{\rm F}+}$ and  $k_{{\rm F}-}$,
  is given by
\begin{eqnarray}
\Delta n &\equiv& 2 \Delta k \, \alpha \nonumber \\
         &\equiv&  2 ( k_{{\rm F}+} - k_{{\rm F}-}) \alpha 
          + \frac{T}{L} \int \d x \, \d \tau 
            \lan 
            \widetilde{k}_{{\rm F}+} - \widetilde{k}_{{\rm F}-}
            \ran  \alpha    \virg  \nonumber \\
\end{eqnarray}
where
$
(\widetilde{k}_{{\rm F}+} - \widetilde{k}_{{\rm F}-})/(2\pi /L) 
      \times 2
\equiv 1/(2L) \sum_{p,\sigma,\mu} \mu \, \rho_{p,\sigma,\mu}
$
and
$\rho_{p,\sigma,\mu}(x,\tau) 
= \psi_{p,\sigma,\mu}^\dagger(x,\tau) \, \psi_{p,\sigma,\mu}(x,\tau)$.
By using the phase variable, 
$
          \partial_x \, \thetatilde _+ (x,\tau) =
          (\pi/\sqrt{2}) 
          \sum_{p,\sigma,\mu} \mu \, \rho_{p,\sigma,\mu} (x,\tau)
$,
the quantity $\Delta n$ is expressed as 
\begin{eqnarray}
\Delta n = - 2 q_0 \alpha + \frac{1}{\sqrt{2}} \frac{T}{L} \alpha
             \int \d x \, \d \tau \, 
             \lan \partial_x \, \thetatilde _+ (x,\tau) \ran \point
    \label{eqn:dn}
\end{eqnarray}
By assuming the scaling relation for Eq. (\ref{eqn:dn}) 
  in a way similar to ref. \cite{Tsuchiizu},
  the renormalization equation for $t(l)$, 
  Eq. (\ref{g0_l}), is derived.

\ecols
\end{document}